\documentclass[amsmath,amssymb,aip,jcp,reprint,floatfix,superscriptaddress]{revtex4-2}

\usepackage{enumitem}
\usepackage{graphicx}
\usepackage{glossaries}
\usepackage[version=4]{mhchem}
\usepackage[per-mode=symbol,separate-uncertainty=true]{siunitx}
\usepackage{hyperref}

\hypersetup{
    unicode,
    colorlinks=true,
    urlcolor=black,
    linkcolor=black,
    citecolor=black
}

\newacronym{cc}{CC}{Clausius-Clapeyron}
\newacronym{dft}{DFT}{density-functional theory}
\newacronym{dof}{DOF}{degrees of freedom}
\newacronym{md}{MD}{molecular dynamics}
\newacronym{mlip}{MLIP}{machine-learned interatomic potential}
\newacronym{nep}{NEP}{neuroevolution potential}
\newacronym{qc}{QC}{quantum corrections}
\newacronym{qha}{QHA}{quasi-harmonic approximation}
\newacronym{rmse}{RMSE}{root mean square error}
\newacronym{scph}{SCPH}{self-consistent phonon}
\newacronym{tdep}{TDEP}{temperature-dependent effective potential}

\makeatletter
\let\oldtheequation\theequation
\renewcommand\tagform@[1]{\maketag@@@{\ignorespaces#1\unskip\@@italiccorr}}
\renewcommand\theequation{(\oldtheequation)}
\makeatother

\DeclareSIUnit\angstrom{\text{Å}}
\DeclareSIUnit\site{\text{site}}
\DeclareSIUnit\atom{\text{atom}}

\newcolumntype{d}{D{.}{.}{-1}}

\newcommand{\phys}{
    Department of Physics and Astronomy,
    Chalmers University of Technology,
    SE 412~96 Gothenburg, Sweden
} 

\newcommand{\llnl}{
    Lawrence Livermore National Laboratory, 
    7000 East Avenue, Livermore, 94550, CA, USA
}

\newcommand{\wise}{
    Wallenberg Initiative Materials Science for Sustainability,
    Chalmers University of Technology,
    41926 Gothenburg, Sweden
}

\begin{document}

\title{Efficient method for calculation of low-temperature phase boundaries}

\author{Lucas Svensson}
\affiliation{\phys}
\affiliation{\wise}

\author{Babak Sadigh}
\email{sadigh1@llnl.gov}
\affiliation{\llnl}

\author{Christine Wu}
\affiliation{\llnl}

\author{Paul Erhart}
\email{erhart@chalmers.se}
\affiliation{\phys}
\affiliation{\wise}

\date{\today}

\begin{abstract}
Understanding phase stability and phase transformations is central to predicting material behavior under varying thermodynamic conditions.
One of the earliest and most influential applications of density functional theory in materials science has been the prediction of pressure-induced phase transitions at \qty{0}{\kelvin}.
Extending these calculations to finite temperatures, however, requires accounting for thermal, quantum, and anharmonic contributions to the free energy, often at significant computational cost.
In this work, we present a general and efficient framework for calculating low-temperature phase boundaries by combining the Clausius-Clapeyron equation with the quasi-harmonic approximation.
This methodology requires a minimal number of calculations, while naturally incorporating internal degrees of freedom as well as quantum and low-order anharmonic effects.
We illustrate the accuracy and efficiency of the approach by constructing the phase diagram of silica in the pressure range from \qtyrange{-2}{12}{\giga \pascal} and temperatures up to \qty{1750}{\kelvin}.
To this end, we employ a machine-learned interatomic potential trained on density functional theory reference data, enabling well-converged free energy estimates via efficient thermodynamic sampling and a rigorous comparison between the proposed framework and free energy integration.
\end{abstract}

\maketitle

\section{Introduction}

Accurate phase diagrams are essential for understanding the behavior of materials under varying thermodynamic conditions.
Predicting phase stability and phase transformations is consequently a cornerstone of materials science, with implications ranging from geology and planetary science to microelectronics and energy storage \cite{Jea.Tho1983, Cop.Smi.Egg.Wan.Ryg.Laz.Haw.Col.Duf2013, Zhe.Sim.Tan.Ke.Nem.Zha.Hu.Lee.Ten.Yan.Wu.Qiu2022, Tia.Jin.Zha.Xu.Wei.Vio.Abr.Yan2017, hainer_2025}.
However, constructing phase diagrams from first principles is often a numerically intensive task, in particular when internal \glspl{dof}, such as lattice distortions or sublattice relaxations, play a significant role in determining phase stability \cite{Sch.Gri2008, Zha.Dal.Wan.Zun2020}.

Electronic structure methods such as \gls{dft} have proven powerful for predicting pressure-induced phase transitions at \qty{0}{\kelvin} by comparing enthalpies across structures \cite{Gra.Bel.And2007}.
However, at finite temperatures, comparing Gibbs free energies across structures over wide ranges of temperatures and pressures can be challenging, as vibrational contributions to both enthalpy and entropy, including zero-point quantum contributions, need to be accounted for.
These contributions are typically evaluated using methods such as the \gls{qha} \cite{Alt.All.Wen.Mor1993} or thermodynamic integration based on \gls{md} simulations \cite{Str.Ber1988, Fre.Ast.de2016}.
Approaches that provide effective harmonic models capturing anharmonic contributions at finite temperatures, such as \gls{tdep} \cite{tdep1} or \gls{scph} methods \cite{TadTsu15, FraRosEri23}, can extend the range of validity of phonon-based free energy calculations to higher temperatures, but at considerably higher computational cost.

Moreover, the presence of soft modes or structural complexity can make phonon calculations particularly delicate, further increasing the computational burden.
These challenges are especially pronounced in materials with complex bonding environments or multiple competing polymorphs.
Silica (\ce{SiO_2}), for example, exhibits a rich array of pressure- and temperature-dependent phases, including tridymite, quartz, coesite, and stishovite, each with distinct structural motifs and vibrational properties \cite{Swa.Sax.Sun.Zha1994}.
Capturing the correct stability fields of such phases requires careful treatment of the vibrational entropy in addition to static lattice energetics.

In this work, we present a general and efficient framework for calculating low-temperature phase boundaries by combining a higher-order expansion of the \gls{cc} equation with a small number of \gls{qha} calculations.
\Gls{qc} are included via a perturbative treatment the terms of which are evaluated using harmonic phonon frequencies within the \gls{qha}.
This accounts for zero-point energy and its effect on the critical pressure at low temperatures, but does not capture anharmonic quantum effects beyond those already implicit in the volume dependence of the \gls{qha} phonon frequencies.
This framework significantly reduces the number of required calculations while naturally incorporating quantum effects and low-order anharmonicity for selected internal \glspl{dof}.

Due to its complex polymorphic landscape and well-studied pressure–temperature behavior, silica serves as an ideal test case for demonstrating our method, which we apply to construct its phase diagram from \qtyrange{-2}{12}{\giga \pascal} and up to \qty{1750}{\kelvin}.
To this end, we train a \gls{mlip} of the \gls{nep} form \cite{FanZenZha21, FanWanYin22} on first-principles \gls{dft} data, and utilize it to perform both the phonon calculations required for our \gls{cc}-\gls{qha}+\gls{qc} framework and free energy integration, which serves as a reference when analyzing the \gls{cc}-\gls{qha}+\gls{qc} results.
This approach enables efficient sampling of the thermodynamic space while retaining ab initio-level accuracy through the underlying training data.

Our results demonstrate the practical advantages of the \gls{cc}-\gls{qha}+\gls{qc} framework for efficiently predicting phase boundaries in materials with complex internal \glspl{dof}, including quantum effects.
Crucially in the present work, the \gls{mlip} serves as a reference method that is sufficiently computationally efficient to enable a one-to-one comparison of the \gls{cc}-\gls{qha}+\gls{qc} approach with free energy integration.
In practice, the \gls{cc}-\gls{qha}+\gls{qc} approach is computationally much less demanding than free energy integration, and naturally accommodates \acrlong{qc}.
Its computational efficiency makes it directly applicable in conjunction with \gls{dft} calculations.

\section{Theory}
\label{sect:theory}

The phase boundary $P^*(T)$ between two phases denoted in the following by superscripts $(1)$ and $(2)$ is determined by the condition that their Gibbs free energies are equal: 
\begin{align*}
G^{\left( 2 \right)}(P^*(T),T) = G^{\left( 1 \right)}(P^*(T),T).
\end{align*}
The Gibbs free energy can be written as 
\begin{align*}
G(P,T)
&= F\left(V(P,T),T\right) + P\,V(P,T) \\ 
&= E(V,T) - T\,S(V,T) + P\,V(P,T),
\end{align*}
where $V(P,T)$ denotes the dependence of the specific volume of a phase on pressure and temperature, $F(V,T)$ is its Helmholtz free energy, $E(V,T)$ its internal energy, and $S(V,T)$ its entropy.
For brevity in the following, we usually omit the explicit dependence of the specific volume $V$ on pressure and temperature although it is always implied.
The internal energy can be further decomposed into two contributions
\begin{equation}
E(V,T) = U_{\text{int}}(V) + E_{\text{vib}}(V,T),
\end{equation}
where the first term on the right-hand side is the potential energy of the perfect lattice and the second term is the vibrational energy. 

In the following, we first derive the temperature dependence of the transition pressure $P^*(T)$ within classical statistical mechanics.
Classical phase boundaries have finite derivatives $dT/dP^*$ at \qty{0}{\kelvin} and are therefore well described by a low-order Taylor expansion near \qty{0}{\kelvin}.
This is not true for the quantum mechanical phase boundaries, which are vertical in the $T$--$P$ plane at \qty{0}{\kelvin} (i.e., $dT/dP^* \to \infty$).
In \autoref{sect:quantum}, we will show how we can derive the quantum phase boundary from the classical one.

\subsection{Classical statistics}
\label{sect:classic}

In a classical system, at \qty{0}{\kelvin}, the vibrational energy vanishes and the transition pressure $P^*(0)$ satisfies the following equation
\begin{align*}
U^{(1)}_{\text{int}}(V^{(1)}(P^*)) &+ P^*\,V^{(1)}(P^*) = \\
U^{(2)}_{\text{int}}(V^{(2)}(P^*)) &+ P^*\,V^{(2)}(P^*)
\end{align*}
To determine the temperature-dependence of the transition pressure, $P^*(T)$, we carry out a second-order Taylor expansion in temperature,
\begin{align}
    P^*(T) = P^*(0) + \left.  \frac{dP^*}{dT} \right |_{T = 0} T + \left. \frac{1}{2}\frac{d^2P^*}{dT^2}\right |_{T = 0}T^2 + \mathcal{O}(T^3).
    \label{eq:transition-pressure}
\end{align}
The first-order derivative can be obtained via the \gls{cc} equation,
\begin{align}\label{eq:CC}
    \frac{dP^*}{dT} = \frac{\Delta S_\text{cl}}{ \Delta V},
\end{align}
where $\Delta S_\text{cl} = S_\text{cl}^{(2)} - S_\text{cl}^{(1)}$ and $\Delta V = V^{(2)} - V^{(1)}$ are the differences in entropy and specific volume between the two considered phases, respectively.

In the classical \qty{0}{\kelvin} limit, vibrations can be considered harmonic, and the classical vibrational entropy can be written as
\begin{align}\label{eq:entropy}
    S_\text{cl}(V) = - k_B \sum_{i = 1}^{3N} \ln\omega_i(V), 
\end{align}
where $N$ is the number of atoms, $\omega_i(V)$ is the volume-dependent angular frequency of the $i$-th phonon mode, and for brevity, we have omitted a benign temperature-dependent term $3Nk_B \left[1 + \ln \left( 2 \pi k_B T/\hbar \right) \right]$.

To evaluate the slope of the classical phase boundary $dP^*/dT$ at \qty{0}{\kelvin}, $S_\text{cl}$ must be calculated for the two phases using \autoref{eq:entropy} and substituted into \autoref{eq:CC}. 

It is possible to obtain higher-order derivatives of the phase boundary at \qty{0}{\kelvin}.
To this end, we differentiate \autoref{eq:CC} with respect to temperature, and for simplicity neglect the explicit temperature dependence of the entropy:
\begin{align}
    \begin{split}
        \frac{d ^2 P^*}{d T^2} &= \frac{d}{d T} \left( \frac{\Delta S_\text{cl}}{\Delta V} \right) \\
        &= \frac{1}{\Delta V} \left[ \frac{\partial V^{\left( 2 \right)}}{\partial T}\left( \frac{\partial S_\text{cl}^{\left( 2 \right)}}{\partial V}  -\frac{d P^*}{d T} \right) \right.  \\
        &\quad\qquad - \left. \frac{\partial V^{\left( 1 \right)}}{\partial T}\left( \frac{\partial S_\text{cl}^{\left( 1 \right)}}{\partial V} - \frac{d P^*}{d T} \right)\right].
    \end{split}     
    \label{eq:second_d}
\end{align}
Note that in \autoref{eq:second_d} $S_{\text{cl}}^{\left( 1 \right)}\left(V^{\left( 1 \right)}\right)$, $S_{\text{cl}}^{\left( 2 \right)} \left(V^{\left( 2 \right)}\right)$, and their derivatives are calculated using \autoref{eq:entropy}, which amounts to the classical \gls{qha}.
Within this approximation, the thermal change of volume can be shown (see Appendix) to be
\begin{equation}
\left.\frac{\partial V}{\partial T} \right |_{T=0}= \frac{V}{B}\frac{\partial S_{\text{cl}}}{\partial V},
\label{eq:Appendix}
\end{equation}
where $B$ is the bulk modulus at \qty{0}{\kelvin}. 
Inserting this relation in \autoref{eq:second_d}, the second derivative of the phase boundary at \qty{0}{\kelvin} can be written as
\begin{align}
    \begin{split}
        \frac{d^2 P^*}{dT^2}
        =& \frac{1}{\Delta V}
        \left[
            \frac{V^{(2)}}{B^{(2)}}
            \frac{\partial S_\text{cl}^{(2)}}{\partial V}
            \left(
                \frac{\partial S_\text{cl}^{(2)}}{\partial V}
                -
                \frac{dP^*}{dT}
            \right)
        \right. \\
        &\left.
            \qquad-
            \frac{V^{(1)}}{B^{(1)}}
            \frac{\partial S_\text{cl}^{(1)}}{\partial V}
            \left(
                \frac{\partial S_\text{cl}^{(1)}}{\partial V}
                -
                \frac{dP^*}{dT}
            \right)
        \right].
    \end{split}
    \label{eq:second_derivative}
\end{align}

We now have all the ingredients for calculating the temperature dependence of a phase boundary to second order in the Taylor expansion \autoref{eq:transition-pressure}, with all terms available via zero-temperature phonon calculations.

\subsection{Quantum statistics}
\label{sect:quantum}

At low temperatures, within the \gls{qha}, the quantum-corrected Helmholtz free energy can be written as
\begin{align}
\begin{split}
    F_{\text{qm}}(V,T) =&\, U_{\text{int}}(V) + \sum_{i = 1}^{3N} \frac{1}{2} \hbar \omega_{i}(V) \\ 
    &+ \sum_{i=1}^{3N}k_B T \ln \left( 1 - e^{-\hbar \omega_{i}(V) / k_B T} \right),
\end{split}
\label{eq:Fqm}
\end{align}
where $N$ is the number of atoms and $\omega_i(V)$ is the angular frequency of the $i$-th phonon mode, which is allowed to be volume-dependent.
The quantum vibrational entropy can be calculated via the relation 
\begin{align*}
S_{\text{qm}} = -\frac{\partial F_{\text{qm}}}{\partial T}.
\end{align*}
However, the strongly non-linear temperature-dependence of $S_{\text{qm}}$ makes a low-order Taylor expansion as in \autoref{eq:transition-pressure} ineffective.
A more appropriate perturbative expansion is in $\delta P(T)$ defined as
\begin{equation}
    \delta P(T) = P^*_{\text{qm}}(T) - P^*_{\text{cl}}(T),
\end{equation}
assuming the magnitude of the quantum corrections to the phase boundary is small, which is a reasonable assumption in the majority of cases. 

For brevity and without loss of generality, we omit the explicit temperature dependence of $\delta P$ and the free energies in the following. We thus write the condition for two-phase coexistence as
\begin{equation}
    G^{(1)}_{\text{qm}}(P^*_{\text{cl}}+\delta P) = 
    G^{(2)}_{\text{qm}}(P^*_{\text{cl}}+\delta P).
\end{equation}
A first-order expansion in $\delta P$ yields
\begin{equation}
 G^{(1)}_{\text{qm}}(P^*_{\text{cl}}) + V^{(1)}_{\text{qm}}(P^*_{\text{cl}}) \, \delta P
 =
 G^{(2)}_{\text{qm}}(P^*_{\text{cl}}) + V^{(2)}_{\text{qm}}(P^*_{\text{cl}}) \, \delta P, 
 \label{eq:qm_cond}
\end{equation}
where we have used the thermodynamic relation 
\begin{equation}
\left.\frac{\partial G}{\partial P}\right|_{P^*} = V(P^*).
\end{equation}
Equation~\eqref{eq:qm_cond} can now be solved to obtain a closed expression for $\delta P$
\begin{equation}
P^*_{\text{qm}} = \frac{F^{(2)}_{\text{qm}}\left(V^{(2)}_{\text{qm}}(P^*_{\text{cl}})\right)-F^{(1)}_{\text{qm}}\left(V^{(1)}_{\text{qm}}(P^*_{\text{cl}})\right)}{V^{(1)}_{\text{qm}}(P^*_{\text{cl}})-V^{(2)}_{\text{qm}}(P^*_{\text{cl}})}
\label{eq:finPqm}
\end{equation}

The above equation is straightforward to evaluate once the quantum-corrected specific volumes $V_{\text{qm}}(P^*_{\text{cl}})$ have been calculated. They are solutions to the equation
\begin{equation}
P^*_{\text{cl}} = -\left.\frac{\partial F_{\text{qm}}}{\partial V}\right|_{V_{\text{qm}}}=-\dot{F}_{\text{qm}}(V_{\text{qm}}).
\label{eq:Pqm}
\end{equation}
The above equation can be solved numerically by a simple root finding algorithm. However, it can also be solved by perturbation theory in orders of $\delta V$ defined as
\begin{equation}
\delta V = V_{\text{qm}} - V_{\text{cl}}. 
\end{equation}
To zeroth order in $\delta V$, \autoref{eq:finPqm} becomes
\begin{equation}
    P^*_{\text{qm}} = \frac{F^{(2)}_{\text{qm}}(V^{(2)}_{\text{cl}})-F^{(1)}_{\text{qm}}(V^{(1)}_{\text{cl}})}{V^{(1)}_{\text{cl}}-V^{(2)}_{\text{cl}}}.
    \label{eq:pressure_QM}
\end{equation}
This level of approximation often suffices for description of the small quantum corrections that are found in common materials.
However, for completeness, we conclude by showing how to derive higher-order terms in perturbation theory.
We therefore rewrite \autoref{eq:Pqm} 
\begin{equation}
   P^*_{\text{cl}} = -\dot{F}_{\text{qm}}(V_{\text{cl}}+\delta V).
\end{equation}
To first order in $\delta V$, the above equation can be solved to yield
\begin{equation}
    \delta V = -\frac{P^*_{\text{cl}}+\dot{F}_{\text{qm}}(V_{\text{cl}})}{\ddot{F}_{\text{qm}}(V_{\text{cl}})},
\end{equation}
where $\ddot{F}_{\text{qm}}$ is the second derivative of the Helmholtz free energy \autoref{eq:Fqm} with respect to volume. Note that the evaluation of $\ddot{F}_{\text{qm}}$ requires calculation of the second derivatives of the phonon frequencies with respect to volume. Inserting now $V_{\text{qm}}=V_{\text{cl}}+\delta V$ in \autoref{eq:finPqm}, we can obtain $P^*_{\text{qm}}$ to a higher order approximation.  

To apply the \gls{cc}-\gls{qha}+\gls{qc} framework, i.e., to evaluate \autoref{eq:transition-pressure} with additional quantum corrections from \autoref{eq:pressure_QM}, we require the following quantities:
\label{sect:workflow}
\begin{enumerate}[label=\Roman*., ref=\Roman*]
    \item \label{item:pc0} the transition pressure at zero temperature, $P^*_{cl}(0)$
    \item \label{item:entropy} the entropy and volume differences between the two phases to evaluate \autoref{eq:CC}
    \item \label{item:bulkmodulus} the derivative of the entropy with respect to volume and the bulk modulus at the transition pressure to calculate \autoref{eq:second_derivative}
    \item \label{item:quantum} the quantum mechanical free energy difference as a function of classical volume, to account for quantum effects using \autoref{eq:pressure_QM}.
\end{enumerate}

While we use volume as the varying \gls{qha}-\gls{dof} in this work, the \gls{cc}-\gls{qha}+\gls{qc} framework is general and may be extended to any internal \glspl{dof}.
This includes, for example, symmetry-preserving lattice distortions or sublattice relaxations where the free-energy landscape can be parameterized as a function of such coordinates.

In summary, the \gls{cc}-\gls{qha}+\gls{qc} framework rests on three approximations.
First, the temperature dependence of the transition pressure is captured by a second-order Taylor expansion, which is justified at low temperatures where higher-order anharmonic contributions are small.
Second, thermal expansion and entropy are evaluated within the \gls{qha}, in which phonon frequencies depend on volume but not explicitly on temperature.
Third, quantum corrections to the transition pressure are included perturbatively to first order in $\hbar$, which is accurate when quantum corrections to the equilibrium volume are small.
Together, these approximations enable the phase boundary to be reconstructed from a small number of phonon calculations at \qty{0}{\kelvin}, without recourse to \gls{md} simulations or full free energy surfaces.

\section{Methods}

\subsection*{Interatomic potential and \texorpdfstring{\gls{dft}}{DFT} calculations}

\begin{figure}
    \centering
    \includegraphics{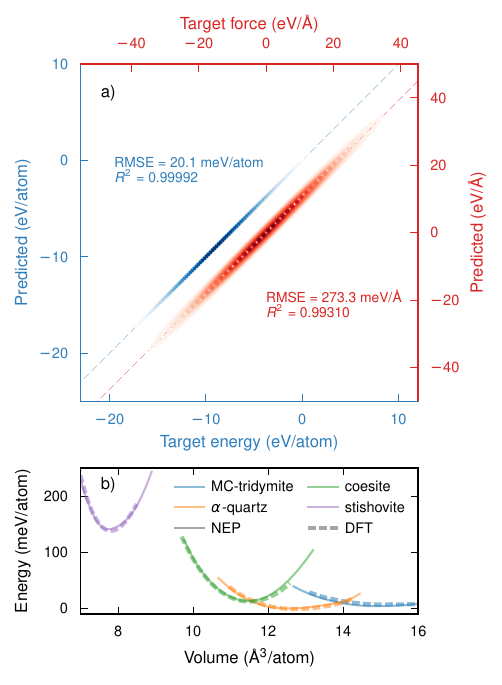}
    \caption{
    (a) Parity plots comparing energies and forces predicted by the \gls{mlip} model constructed in this work against \gls{dft} reference data.
    (b) Energy--volume curves computed using the \gls{mlip} model, shown alongside corresponding results from \gls{dft} calculations for validation.
    }
    \label{fig:nep}
\end{figure}

To demonstrate the \gls{cc}-\gls{qha}+\gls{qc} framework and to provide a reference via free energy integration, we constructed a \gls{mlip} of the \gls{nep} form \cite{FanZenZha21, FanWanYin22, XuBuPan25, LinRahFra24} using an active learning strategy \cite{FraWikErh23} based on \gls{dft} calculations with the r2SCAN exchange-correlation functional \cite{r2scan} (see \autoref{snote:potential} and \autoref{snote:dft} for full details of the model construction \cite{LarMorBlo17, EriFraErh19} and \gls{dft} settings \cite{Blo94, KreJou99, KreHaf93, KreFur1996-1, KreFur1996-2}).
The training set consisted of \num{1218} structures covering crystalline and amorphous configurations of \ce{SiO2} and \ce{Si} across a range of pressures and temperatures.
The final model achieves \glspl{rmse} of \qty{20}{\milli\electronvolt\per\atom} for energies and \qty{273}{\milli\electronvolt\per\angstrom} for forces, with coefficients of determination $R^2 > 0.99$ in both cases (\autoref{fig:nep} and \autoref{snote:potential}).
The energy--volume curves and enthalpy differences predicted by the \gls{nep} model are in very good agreement with the underlying \gls{dft} reference data (see \autoref{sfig:enthalpy-pressure} and \autoref{sfig:energy-volume}), confirming that the model faithfully reproduces the energy landscape relevant for the phase boundaries studied here.
We note that the level of agreement with experiment \cite{Pol14} observed below should be attributed primarily to the r2SCAN functional used in the training data.
The \gls{mlip} models and \gls{dft} reference data are available on Zenodo at \url{https://doi.org/10.5281/zenodo.14925353}.

\subsection*{Clausius-Clapeyron relation and quasi-harmonic approximation}

To efficiently compute phase boundaries at low temperatures, we combine the \gls{cc} equation \ref{eq:CC} with \gls{qha} phonon calculations. 
This approach captures quantum effects and anharmonic corrections while requiring only a limited set of force calculations (\autoref{sect:theory}).
In contrast to conventional \gls{qha}-based phase diagram construction, which requires full free energy surfaces, the \gls{cc}-\gls{qha}+\gls{qc} approach reconstructs phase boundaries from local thermodynamic derivatives, significantly reducing computational cost.

The transition pressure at zero temperature, $P^*(0)$, (see step~\ref{item:pc0} in workflow in \autoref{sect:workflow}) was determined by computing the enthalpy as a function of pressure for each phase and identifying the transition pressure at which their enthalpy curves intersect (see \autoref{sfig:enthalpy-pressure} of the Supplementary Material). 

To compute the volume and entropy differences (step~\ref{item:entropy}), structures of each phase relaxed at $P^*(0)$ were used.
The force constants were determined using the \textsc{hiphive} package \cite{EriFraErh19} and subsequently passed to the \textsc{phonopy} package \cite{togo_first-principles_2023, togo_implementation_2023} to compute the classical entropy of each phase.

The entropy derivative and bulk modulus (step~\ref{item:bulkmodulus}) were evaluated using structures with volumes \num{0.2}\% above and below that of the structures relaxed at $P^*(0)$.
The entropy derivative was obtained through numerical differentiation as
\begin{align*}
    \left. \frac{\partial S}{\partial V} \right|_{V_0} = \frac{S(V_0 + \delta V) - S(V_0 - \delta V)}{2 \delta V}
\end{align*}
and the bulk modulus as
\begin{align*}
    B = -V_0 \left. \frac{dP}{dV} \right|_{V_0}
    = - V_0 \frac{P(V_0 + \delta V) - P(V_0 - \delta V)}{2 \delta V}.
\end{align*}
The classical and quantum mechanical free energy differences between the phases (step~\ref{item:quantum}) were then calculated using the same set of force constants in combination with \textsc{phonopy}.

\subsection*{Free energy integration and molecular dynamics}

To validate the \gls{cc}-\gls{qha}+\gls{qc} framework, the phase diagram was additionally constructed via free energy integration using \gls{md} simulations.
To this end, we performed thermodynamic integration using both adiabatic switching \cite{FreLad84} and reversible scaling \cite{KonAntYip99, Fre.Ast.de2016} as implemented in \textsc{gpumd} \cite{FanWanYin22, XuBuPan25}. 
To obtain the second-order transition boundary between $\alpha$-quartz and $\beta$-quartz, standard \gls{md} simulations were conducted at various pressures (\autoref{sfig:alpha-quartz-sims} and \autoref{sfig:temperature-dependence} in the Supplementary Material).
The heat capacities exhibit a pronounced peak at the $\alpha$-$\beta$-quartz transition temperature (\autoref{sfig:alpha-quartz-sims}b), which yields the phase boundary as a function of pressure.

\section{Results}

\begin{figure*}
    \centering
    \includegraphics{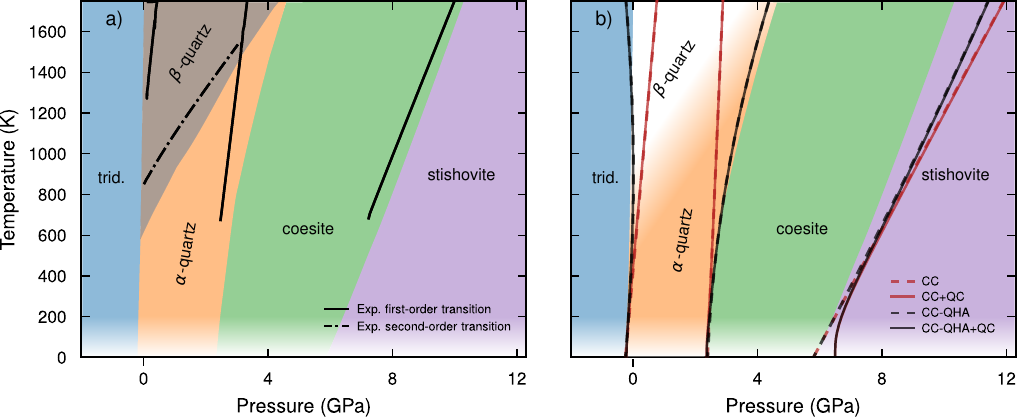}
    \caption{
    Temperature--pressure phase diagram of \ce{SiO2}.
    Colored regions indicate the phase diagram predicted using free energy integration based on the \gls{nep} model.
    The faded low-temperature region marks the lowest temperature at which free energy integration remains accurate.
    (a) Phase boundaries adapted from experimental data reported in Ref.~\citenum{Swa.Sax.Sun.Zha1994}.
    (b) Phase boundaries obtained using first-order expansion without \gls{qc} (\gls{cc}), first-order expansion with \gls{qc} (\gls{cc}+\gls{qc}), second-order expansion without \gls{qc} (\gls{cc}-\gls{qha}), and second-order expansion with \gls{qc} (\gls{cc}-\gls{qha}+\gls{qc}).
    }
    \label{fig:phase-diagram}
\end{figure*}

\begin{figure}
    \centering
    \includegraphics{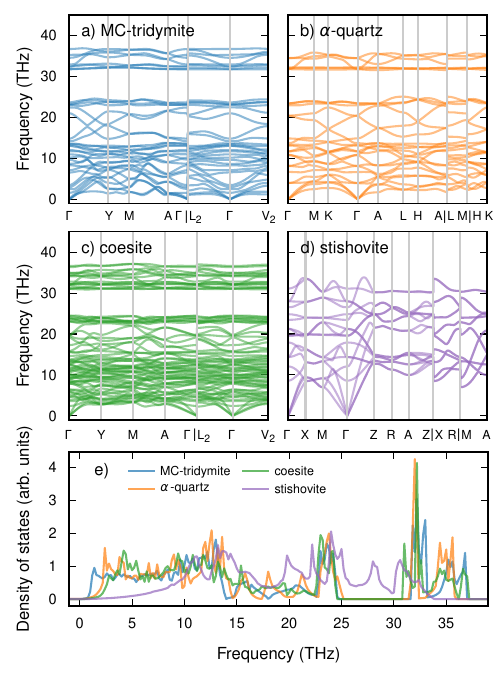}
    \caption{
        Phonon dispersion relations of (a) tridymite, (b) $\alpha$-quartz, (c) coesite, and (d) stishovite, along with (e) the corresponding phonon densities of states for all four structures.
    }
    \label{fig:phonons}
\end{figure}

Figure \ref{fig:phase-diagram} compares phase diagrams for \ce{SiO2} obtained using the CALPHAD approach based on experimental data \cite{Swa.Sax.Sun.Zha1994} as well as using the \gls{nep} model via free energy integration and the \gls{cc}-\gls{qha}+\gls{qc} framework introduced here, respectively.
We first observe that the \gls{nep} model predicts a phase diagram in good agreement with experiment (\autoref{fig:phase-diagram}a).
The low-pressure transition between tridymite and $\alpha$-quartz, as well as the high-pressure transition between coesite and stishovite, are found to be in near-quantitative agreement.
However, slight deviations are observed for the transition between $\alpha$-quartz and coesite, resulting in a modest overstabilization of the $\alpha$-quartz and $\beta$-quartz phases.
Both the agreement and the remaining deviations should be attributed to the underlying exchange-correlation functional.
This assessment is supported by the observation that the \gls{mlip} constructed in Ref.~\citenum{Erh.Roh.Alb.Der2024} using a different \gls{mlip} format and training set, albeit using a closely related exchange-correlation functional, predicts a phase diagram that resembles the one obtained via the \gls{nep} model here.

Since the phonon calculations used to predict the phase diagram within our \gls{cc}-\gls{qha}+\gls{qc} framework (\autoref{fig:phase-diagram}b) are based on the \gls{nep} model, the phase diagram obtained via free energy integration serves as a natural reference for assessing the accuracy of our approach.
This comparison shows that the \gls{cc}-\gls{qha}+\gls{qc} framework predicts the phase boundaries between tridymite and $\alpha$-quartz as well as between $\alpha$-quartz and coesite in excellent agreement with the reference phase diagram.
With respect to the stishovite--coesite transition, the \gls{cc}-\gls{qha}+\gls{qc} approach, however, yields a boundary with a somewhat flatter slope.
It should be noted that the boundary between $\alpha$- and $\beta$-quartz marks a continuous transition that is not numerically accessible from free energies, and therefore cannot be captured by the \gls{cc}-\gls{qha}+\gls{qc} approach.

The accuracy of the \gls{cc}-\gls{qha}+\gls{qc} framework is bounded by the applicability of the \gls{qha} and the low-temperature Taylor expansion; the method is not expected to remain accurate when anharmonic contributions become significant.
For the tridymite--quartz and quartz--coesite boundaries, the framework remains in excellent agreement with the free energy integration reference up to the highest temperatures considered here ($\sim\!\qty{1500}{\kelvin}$).
In the case of the coesite--stishovite boundary, deviations become noticeable already around \qty{600}{\kelvin}, which we attribute to the unusually large contrast in vibrational stiffness between the two phases, evident from both the phonon dispersions (\autoref{fig:phonons}) and the elastic constants (\autoref{stab:SiO2-elastic-stiffness-tensors}), that amplifies anharmonic contributions.
Methods that provide effective harmonic models at finite temperatures, such as \gls{tdep} \cite{tdep1} or \gls{scph} approaches \cite{TadTsu15, FraRosEri23}, could in principle supply improved force constants as input to the present framework, potentially extending its range of applicability to higher temperatures.

It is also apparent that the \gls{cc}-\gls{qha}+\gls{qc} approach yields a noticeable and consistent improvement compared to the classical \gls{cc} relation (red lines in \autoref{fig:phase-diagram}b).
This effect is most evident for the $\alpha$-quartz-coesite and tridymite-$\alpha$-quartz boundaries, for the latter of which the \gls{cc} relation even yields the wrong slope.

Examining the low-temperature region of the coesite--stishovite boundary (\autoref{fig:phase-diagram}b) highlights the importance of incorporating quantum effects within the \gls{cc}-\gls{qha}+\gls{qc} framework.
When quantum effects are neglected, the boundary exhibits a finite slope $dT/dP^*$ at \qty{0}{\kelvin}, whereas the inclusion of quantum effects correctly, and as expected from thermodynamic principles, drives this slope to infinity (i.e., the boundary becomes vertical in the $T$--$P$ plane), consistent with the vanishing of the quantum mechanical entropies of both phases at \qty{0}{\kelvin} as dictated by \autoref{eq:CC}.

The phonon dispersions (\autoref{fig:phonons}) reveal that stishovite exhibits significantly fewer low-frequency phonon modes compared to coesite, indicating a higher vibrational stiffness, which is also visible in the elastic constants (\autoref{stab:SiO2-elastic-stiffness-tensors}).
This stiffer character leads to a smaller vibrational entropy at low temperatures. 
Consequently, the larger entropy difference between the two phases leads to a stronger temperature dependence of the transition pressure, further highlighting the relevance of quantum effects in accurately describing this phase boundary.

\section{Conclusions}

We have introduced a framework that combines the \acrlong{cc} equation with the \acrlong{qha} and \acrlong{qc} (\gls{cc}-\gls{qha}+\gls{qc}) and demonstrated that it provides an efficient and accurate method for determining phase boundaries in solid-phase systems.
In particular, we have shown that this approach captures the thermodynamic behavior of \ce{SiO2} polymorphs across a wide range of pressures and temperatures.

Compared to free energy integration methods based on \gls{md}, the \gls{cc}-\gls{qha}+\gls{qc} framework reduces the computational cost substantially while yielding comparable accuracy.
In principle, only six phonon calculations, three per phase, are required to map out a two-phase boundary, making the method directly applicable with \gls{dft} calculations.
Free energy integration, by contrast, requires extensive \gls{md} sampling at each state point, making it impractical to perform directly with \gls{dft}; the use of \glspl{mlip} makes this task more tractable but requires prior model construction.
Moreover, incorporating quantum nuclear effects in free energy integration requires path-integral \gls{md} simulations, increasing the computational effort by roughly an order of magnitude.
The \gls{cc}-\gls{qha}+\gls{qc} approach therefore strikes a balance between efficiency, accuracy, and natural incorporation of quantum effects, making it well suited for applications that require \gls{dft}-level accuracy or high-throughput phase boundary calculations.

Importantly, and as expected from thermodynamic principles, the inclusion of quantum vibrational effects within the \gls{cc}-\gls{qha}+\gls{qc} framework correctly reproduces the infinite slope of phase boundaries at zero temperature.
This is especially relevant for systems where the vibrational entropy differs strongly between neighboring phases (exemplified by the coesite--stishovite boundary), influencing the shape and location of the phase boundaries.
Crucially, this is achieved without recourse to path-integral \gls{md} simulations, by incorporating the quantum correction via the zero-point free energy difference at the \gls{qha} level.

The accuracy of the framework is bounded by the validity of the \gls{qha} and the low-temperature Taylor expansion.
In the present application to silica, excellent agreement with the free energy integration reference is obtained up to $\sim\!\qty{1500}{\kelvin}$ for the tridymite--$\alpha$-quartz and $\alpha$-quartz--coesite boundaries.
For the coesite--stishovite boundary, deviations become apparent around \qty{600}{\kelvin}, likely due to the large stiffness contrast between those phases.
Replacing the \gls{qha} force constants with temperature-dependent ones from \gls{tdep} \cite{tdep1} or \gls{scph} \cite{TadTsu15, FraRosEri23} could extend the approach to more strongly anharmonic systems or higher temperatures.

Although volume is the only quasi-harmonic \gls{dof} considered in this work, the \gls{cc}-\gls{qha}+\gls{qc} framework is generalizable to any number of internal coordinates that can be treated within the \gls{qha}.

\appendix

\section{Relationship for thermal expansion coefficient}

In this section, we show that within the \gls{qha} approximation, \autoref{eq:Appendix} is satisfied. We start with the expression for the pressure within \gls{qha}
\begin{equation}
    P(T) = -\frac{\partial U_{\text{int}}}{\partial V} + T\frac{\partial S_{\text{cl}}}{\partial V},
\end{equation}
where $U_{\text{int}}$ and $S_{\text{cl}}$ are defined in Sect.~\ref{sect:classic}. Now, under isobaric conditions, the temperature dependence of the specific volume follows from the constant-pressure condition. At $0$~K, we thus have
\begin{equation}
0 = \left.\frac{dP}{dT}\right|_{T=0} = -\frac{\partial^2 U_{\text{int}}}{\partial V^2}\left.\frac{\partial V}{\partial T} \right|_{T=0
}+ \frac{\partial S_{\text{cl}}}{\partial V}.
\end{equation}
Solving the above equation, using the relation 
\begin{equation}
\frac{B}{V} = \frac{\partial^2 U_{\text{int}}}{\partial V^2},
\end{equation}
we obtain \autoref{eq:Appendix}.

\section*{Supplementary Material}

The supplementary material provides further details on the training of the \gls{nep} model and the \gls{dft} calculations used to generate the training data, as well as benchmark results, including parity plots, energy-volume curves, thermal expansion, and phonon dispersions.

\section*{Acknowledgments}

This work was supported by the Swedish Research Council (Nos. 2020-04935, 2021-05072, and 2025-03999) and the Wallenberg Initiative Materials Science for Sustainability.
Part of this work was performed under the auspices of the US Department of Energy by Lawrence Livermore National Laboratory under Contract DE-AC52-07NA27344.
Some of the computations were enabled by resources provided by the National Academic Infrastructure for Supercomputing in Sweden (NAISS) at PDC, C3SE, and NSC, partially funded by the Swedish Research Council through grant agreement no. 2022-06725 as well as the Berzelius resource provided by the Knut and Alice Wallenberg Foundation at NSC.

\section*{Conflict of Interest}

The authors have no conflicts to disclose.

\section*{Data availability}

The \gls{mlip} model, including the model ensemble constructed in this work as well as the \gls{dft} reference data used for its construction, are available on Zenodo at \url{https://doi.org/10.5281/zenodo.14925353}.


%

\end{document}